\documentstyle[preprint,tighten,aps]{revtex}
\begin{document}
\draft
\baselineskip=0.95\baselineskip
\title
{Renormalization group and nonlinear susceptibilities of 
cubic ferromagnets at criticality}

\author{D.~V.~Pakhnin$^1$ and A.~I.~Sokolov$^{1,2}$}

\address
{$^1$Department of Physical Electronics, 
Saint Petersburg Electrotechnical University, \\ Professor Popov Street 5,
Saint Petersburg 197376, Russia, \\
$^2$Department of Physics, Saint Petersburg Electrotechnical University, \\
Professor Popov Street 5, Saint Petersburg 197376, Russia}

\maketitle

\begin{abstract}
For the three-dimensional cubic model, the nonlinear susceptibilities of 
the fourth, sixth, and eighth orders are analyzed and the parameters 
$\delta^{(i)}$ characterizing their reduced anisotropy are evaluated at 
the cubic fixed point. In the course of this study, the renormalized 
sextic coupling constants entering the small-field equation of state are 
calculated in the four-loop approximation and the universal values of 
these couplings are estimated by means of the Pad\'e-Borel-Leroy 
resummation of the series obtained. The anisotropy parameters are found 
to be: $\delta^{(4)} = 0.054 \pm 0.012$, $\delta^{(6)} = 0.102 \pm 0.02$, 
and $\delta^{(8)} = 0.144 \pm 0.04$, indicating that the anisotropic 
(cubic) critical behavior predicted by the advanced higher-order 
renormalization-group analysis should be, in principle, visible in 
physical and computer experiments.  

\vspace{1.5cm}
 

\end{abstract}

\section{Introduction}
\label{sec:1}

Three decades ago Wilson and Fisher discovered that in the vicinity of 
the critical point the order parameter fluctuations may change the 
effective anisotropy of the system drastically \cite{WF}. Having studied 
the anisotropic XY model within the $\epsilon$ expansion they found that 
approaching $T_c$ this model either becomes effectively isotropic or 
developes its anisotropy further \cite {WF} until the fluctuation-induced 
first-order phase transition takes place \cite {W,KW,LP}. What scenario 
the system chooses depends on the value and sign of the anisotropic 
coupling constant in the initial Landau-Wilson Hamiltonian. In the 
course of studying of the generalized cubic model, the crucial role of 
the order parameter dimensionality $n$ was revealed: for $n < n_c$ the 
model undergoing the continuous phase transition demonstrates 
isotropic critical behavior while for $n > n_c$ it remains anisotropic 
at criticality \cite{A}. The numerical value of the marginal spin 
dimensionality $n_c$ separating these two regimes is of prime physical 
importance since it fixes the true mode of the critical behavior of 
real cubic ferromagnets and of some other systems of interest. 

First estimates of $n_c$ for the cubic model were obtained within 
the lower-order renormalization-group (RG) calculations \cite{KW,S77,NR} 
and turned out to be in favor of the inequality $n_c > 3$ indicating  
that cubic ferromagnets should belong to the class of universality of 
the three-dimensional (3D) Heisenberg model. The classical technique 
of the high-temperature expansions seemed to support this conclusion 
\cite{FVC}. However, the resummation of the three-loop RG expansions 
\cite{MSI,Sh89} and subsequent higher-order RG calculations performed 
both in three \cite{MSS,PS,CPV} and $(4 - \epsilon)$ (Ref.\cite{KS}) 
dimensions shifted the numerical estimate for $n_c$ downwards, fixing 
it below 3 \cite{MSI,Sh89,MSS,PS,CPV,KS,KT,KTS,SAS,FHY1,FHY2,V,PVr}. 
Remarkable consensus between different field-theoretical approaches and 
resummation techniques was achieved in the course of this study: 3D RG 
calculations \cite{CPV}, resummed $\epsilon$ expansion \cite{SAS}, biased 
$\epsilon$ expansion technique \cite{CPV}, and pseudo-$\epsilon$ 
expansion analysis \cite{FHY2} lead to $n_c = 2.89$, $n_c = 2.855$, 
$n_c = 2.87$, and $n_c = 2.862$, respectively. Moreover, the predicted 
numerical value of $n_c$ was found to be stable with respect to the 
order of perturbation theory: starting from the four-loop approximation, 
the 3D RG estimate stays practically unchanged varying around 2.9 within 
the interval $\pm 0.02$ \cite{MSS,PS,CPV}. Hence, according to the 
up to date, most accurate theoretical data, cubic ferromagnets should 
demonstrate the specific -- cubic -- critical behavior with a special 
set of critical exponents.

Since $n_c$ is very close to the physical value $n = 3$, 
the fixed point governing the cubic critical behavior lies very near 
the Heisenberg fixed point at the RG flow diagram and the critical 
exponents for both fixed points almost coincide. For instance, the 
susceptibility exponent $\gamma$ is equal to 1.3895(50) for the 3D  
Heisenberg model \cite{GZ1} while at the cubic fixed point  
$\gamma = 1.390(6)$ \cite{CPV}. It is clear that measuring the  
critical exponents one cannot distinguish between the cubic and 
Heisenberg critical behaviors. 

In such a situation, some alternative physical quantities should be 
addressed to clear up how the systems with the cubic symmetry 
behave approaching $T_c$. In this paper, the nonlinear susceptibilities 
of the cubic model are studied in the framework of the field-theoretical 
RG approach in three dimensions, with particular attention paid to 
their anisotropy at criticality. It will be shown that 
the anisotropy of nonlinear susceptibilities is sensitive to the type  
of critical asymptotics and its measurement can be used, in principle, 
for detection of the cubic (non-Heisenberg) critical behavior, provided 
the anisotropy is as strong as the modern higher-order RG calculations 
predict.

\section{Nonlinear susceptibilities and effective coupling constants}
\label{sec:2}

In the critical region, the expansion of the free energy of the 
cubic model in powers of the magnetization components $M_{\alpha}$ 
may be written in the form:   
\begin{eqnarray}   
F(M_{\alpha}, m) &=& F(0, m) +  {\frac{1}{2}} m^{2 - \eta} 
M_{\alpha}^2 + m^{1 - 2\eta}(u_4 + v_4 \delta_{\alpha \beta}) 
M_{\alpha}^2 M_{\beta}^2 
\nonumber\\&&
+ m^{- 3\eta}(u_6 + q_6 \delta_{\alpha \beta} + 
v_6 \delta_{\alpha \beta} \delta_{\alpha \gamma}) 
M_{\alpha}^2 M_{\beta}^2 M_{\gamma}^2 + ...
\label{eq:1} 
\end{eqnarray}
where $\eta$ is a Fisher exponent, $m$ being an inverse correlation 
length, and $u_4$, $v_4$, $u_6$, $q_6$, and $v_6$ are dimensionless 
effective coupling constants acquiring, under $T \to T_c$, certain 
universal values. These coupling constants are related to the 
nonlinear susceptibilities defined in a conventional way: 
\begin{equation}
\chi_{\alpha \beta \gamma \delta}^{(4)} = 
{\frac{\partial^3 M_{\alpha}}{{\partial H_{\beta}}{\partial H_{\gamma}}
{\partial H_{\delta}}}} \Bigg\arrowvert_{H = 0}, 
\qquad \quad 
\chi_{\alpha \beta \gamma \delta \mu \nu}^{(6)} = 
{\frac{\partial^5 M_{\alpha}}{{\partial H_{\beta}}{\partial H_{\gamma}}
{\partial H_{\delta}}{\partial H_{\mu}}{\partial H_{\nu}}}} 
\Bigg\arrowvert_{H = 0}.
\label{eq:2} \\
\end{equation}
Of prime importance are the nonlinear susceptibilities in two particular 
cases, when (i) an external magnetic field is parallel to a cubic axis 
($\chi_{c}^{(i)}$) and (ii) it is oriented along a cell space diagonal 
($\chi_{d}^{(i)}$). For these two highly symmetric directions the 
differences between the corresponding values of the nonlinear 
susceptibilities are known to be maximal: i. e., the anisotropy is most 
pronounced. It is easy to show that  
\begin{equation}
\chi_{c}^{(4)} = - 24 {\frac{\chi^2}{m^3}} (u_4 + v_4), 
\qquad \quad 
\chi_{d}^{(4)} = - 24 {\frac{\chi^2}{m^3}} \Biggl(u_4 + 
{\frac{v_4}{3}} \Biggr).
\label{eq:3} \\
\end{equation}
\begin{eqnarray}
\chi_{c}^{(6)} = 720 {\frac{\chi^3}{m^6}} \Biggl[8 
(u_4 + v_4)^2 - u_6 - q_6 - v_6 \Biggr],  
\nonumber \\ 
\chi_{d}^{(6)} = 720 {\frac{\chi^3}{m^6}} \Biggl[8 \Biggl(u_4 + 
{\frac{v_4}{3}} \Biggr)^2 - u_6 - {\frac{q_6}{3}} - 
{\frac{v_6}{9}} \Biggr].
\label{eq:4} 
\end{eqnarray}
where $\chi$ is a linear susceptibility. To characterize the anisotropy 
strength, we define the reduced parameters 
\begin{equation}
\delta^{(4)} = {\frac{|\chi_{c}^{(4)} - 
\chi_{d}^{(4)}|}{\chi_{c}^{(4)}}},
\qquad \qquad 
\delta^{(6)} = {\frac{|\chi_{c}^{(6)} - \chi_{d}^{(6)}|} 
{\chi_{c}^{(6)}}}.
\label{eq:5} 
\end{equation}
Below, they will be estimated at criticality. 

Let us start from the lower-order nonlinear susceptibility 
$\chi^{(4)}$. The coordinates of the cubic fixed 
point $u_4^*$ and $v_4^*$ in three dimensions are known from the 
higher-order RG calculations, with resummed four-, five-, and six-loop 
RG expansions yielding very close numerical results \cite{MSS,CPV}. 
Considering recent six-loop RG estimates as the most reliable ones, 
we accept $u_4^* = 0.755 \pm 0.010$, $v_4^* = 0.067 \pm 0.014$ 
\cite{CPV}. The substitution of these numbers into Eqs. (3) and (5) 
gives 
\begin{equation}
\delta^{(4)} = 0.054 \pm 0.012.
\label{eq:6} \\
\end{equation}
In the next section, the universal critical values of the sextic coupling 
constants will be evaluated and an estimate for $\delta^{(6)}$ will be 
obtained.

\section{RG expansions for sextic couplings and anisotropy of  
sixth-order susceptibility}
\label{sec:3}

Let us proceed to determination of $\chi_{c}^{(6)}$, 
$\chi_{d}^{(6)}$, and $\delta^{(6)}$. As seen from Eq. (4), we are in 
a position to calculate the effective coupling constants $u_6$, $q_6$, 
and $v_6$. They will be found perturbatively, using the 
field-theoretical RG approach in three dimensions. Our analysis  
is based on the well-known Landau-Wilson Hamiltonian of the 
3D $n$-vector cubic model:
\begin{equation}
H = {\frac{1}{2}} 
\int d^3x \Biggl[ m_0^2 \varphi_{\alpha}^2 + 
(\nabla \varphi_{\alpha})^2 
+ {\frac{u_0}{12}} \varphi_{\alpha}^2 \varphi_{\beta}^2 
+ {\frac{v_0}{12}} \varphi_{\alpha}^4 \Biggr], 
\label{eq:7}
\end{equation}
where $m_o^2$ is the reduced deviation from the mean-field transition 
temperature. All the RG calculations are carried out within a massive 
theory under zero-momentum normalizing conditions for the renormalized 
Green function $G_R(p,m)$ and the four-point 
vertices $U_R(p_i,m,u,v,)$, $V_R(p_i,m,u,v,)$:
\begin{eqnarray}
G_R^{-1}(0,m) = m^2, \ \ \ \  
{\frac{\partial G_R^{-1}(p,m)}{\partial p^2}} 
\Big\arrowvert_{p^2 = 0} = 1, \ \ \ \
\nonumber \\
U_R(0,m,u,v) = m u, \ \ \ \ \ V_R(0,m,u,v) = m v. \
\label{eq:8}
\end{eqnarray}
Here, the value of the one-loop vertex graph including the factor 
$(n + 8)$ is absorbed in $u$ and $v$ in order to make the coefficient 
for the $u^2$ term in the $\beta_u$ function equal to unity. The quartic 
effective couplings $u$ and $v$ thus defined are related to $u_4$ and 
$v_4$ entering Eqs. (1), (3), and (4) in an obvious way: 
\begin{equation}
u = {\frac{n + 8}{2 \pi}} u_4, 
\qquad v = {\frac{n + 8}{2 \pi}} v_4. 
\label{eq:9}
\end{equation}

We limit ourselves to the four-loop RG approximation that proved to 
lead to quite a good numerical estimate for the universal value of the 
sextic effective coupling for the 3D O(n)-symmetric model \cite{S,SOUK}.
Since the symmetry factors and integrals for all the relevant Feynman 
diagrams have been found earlier \cite{SOUK}, what we have to do 
here is the calculation of the tensor (field) factors for the six-point 
vertex graphs and mass insertions generated by the O(n)-symmetric and 
cubic interactions. Performing this calculation and then renormalizing 
the "bare" perturbative expansions for the six-point vertices precisely 
in the same way as has been done for the isotropic n-vector model 
\cite{S,SOUK}, we arrive at the following four-loop RG expansions: 

\begin{eqnarray}
u_6 &=& {8 \pi^2 \over 3} {u^2 \over (n+8)^3} 
\Biggl[(n + 26) u + 9 v - {1 \over (n + 8)} \biggl({{34 n 
+ 452} \over 3} u^2 + 124 u v + 18 v^2 \biggr) 
\nonumber\\&&
+~{1 \over (n + 8)^2} \biggl[ (1.065025 n^2 + 157.42454 n 
+ 1323.0960) u^3 
\nonumber\\&&
+~(19.382741 n + 1873.9825) u^2 v 
+ (2.3101121 n + 679.33934) u v^2 + 100.90652 v^3 \biggr] 
\nonumber\\&&
-~{1 \over (n + 8)^3} \biggl[ (- 0.0638 n^3 + 52.451 n^2 
+ 2314.99 n + 14387.6) u^4 
\nonumber\\&&
+~(- 2.3434 n^2 + 1006.56 n + 29601.0) u^3 v 
+ (100.156 n + 18985.4) u^2 v^2 
\nonumber\\&&
+~(11.286 n + 5872.61) u v^3 
+ 782.865 v^4 \biggr] \Biggr].
\label{eq:10}
\end{eqnarray}

\begin{eqnarray}
q_6 &=& {8 \pi^2 \over 3} {u v \over (n+8)^3} \Biggl[ 72 u + 27 v 
- {1 \over (n+8)} \biggl((4 n + 520) u^2 + 468 u v + 108 v^2 \biggr) 
\nonumber\\&&
+~{1 \over (n+8)^2} \biggl[(-1.1357879 n^2 + 149.54004 n 
+ 5366.1581) u^3 
\nonumber\\&&
+~(- 32.454138 n + 8267.9231) u^2 v + 4107.2194 u v^2 
+ 775.79040 v^3 \biggr] 
\nonumber\\&&
-~{1 \over (n+8)^3} \biggl[(0.3986 n^3 - 19.2521 n^2 + 3351.86 n 
+ 66591.7) u^4 
\nonumber\\&&
+~ (11.6195 n^2 + 115.223 n + 148119) u^3 v 
+ (- 419.561 n + 119322) u^2 v^2 
\nonumber\\&&
+~46560.3 u v^3 + 7520.89 v^4 
\biggr] \Biggr].
\label{eq:11}
\end{eqnarray}

\begin{eqnarray}
v_6 &=& {8 \pi^2 \over 3} {v^2 \over (n+8)^3} \Biggl[54 u + 27 v 
- {1 \over (n+8)} \biggl[(- 6 n + 492) u^2 + 540 u v + 162 v^2 \biggr] 
\nonumber\\&&
+~{1 \over (n+8)^2} \biggl[(1.3828190 n^2 - 50.171743 n 
+ 5947.5257) u^3 
\nonumber\\&&
+~(- 88.612482 n + 10696.342) u^2 v 
+ 6632.1372 u v^2 + 1481.5855 v^3 \biggr] 
\nonumber\\&&
-~{1 \over (n+8)^3} \biggl[(-0.4873 n^3 + 6.2132 n^2 + 114.760 n 
+ 83872.7) u^4 
\nonumber\\&&
+~(25.5088 n^2 - 2440.17 n + 212729) u^3 v 
+ (- 1281.62 n + 205263) u^2 v^2 
\nonumber\\&&
+~93009.1 u v^3 + 16755.0 v^4 \biggr] \Biggr].
\label{eq:12}
\end{eqnarray}

These RG expansions should obey some exact relations appropriate to 
systems with cubic anisotropy. Indeed, for $n = 2$ the model 
(7) is known to possess a specific symmetry property. If the field 
$\varphi_{\alpha}$ undergoes the transformation 
\begin{equation}
\varphi_1 \to {\frac{\varphi_1 + \varphi_2}{\sqrt 2}} \ \ , \quad 
\varphi_2 \to {\frac{\varphi_1 - \varphi_2}{\sqrt 2}} \ \ , 
\label{eq:13}
\end{equation}
the quartic coupling constants are also transformed,
\begin{equation}
u \to u + {\frac{3}{2}} v \ \ , \quad v \to -v \ \ , 
\label{eq:14}
\end{equation}
but the structure of the Hamiltonian itself remains unchanged 
\cite{WF}. Since the RG functions of the problem are completely 
determined by the structure of the Hamiltonian, the RG equations 
should be invariant with respect to any transformation conserving 
this structure \cite{K}. This should be also true for all the 
expressions relating various universal quantities to each other. 
If, under $n = 2$, we apply the transformation (13) to the 
magnetization components in Eq. (1), the free-energy expansion 
remains the same, provided $u_4$, $v_4$ are replaced according to 
Eq. (14) while the sextic coupling constants are transformed as 
prescribed below:  
\begin{equation}
u_6 + {\frac{q_6}{2}} + {\frac{v_6}{4}} \to 
u_6 + q_6 + v_6, \ \
\quad {\frac{2}{3}}q_6 + v_6 \to - {\frac{2}{3}}q_6 - v_6. 
\label{eq:15}
\end{equation}
It means that the following relations between the effective coupling 
constants should hold:    
\begin{eqnarray}
u_6(u, v) + {\frac{q_6(u, v)}{2}} + {\frac{v_6(u, v)}{4}} &=& 
u_6(u + {\frac{3}{2}}v, -v) + q_6(u + {\frac{3}{2}}v, -v) 
+ v_6(u + {\frac{3}{2}}v, -v), 
\nonumber\\
{\frac{2}{3}}q_6(u, v) + v_6(u, v) &=& - {\frac{2}{3}}q_6(u 
+ {\frac{3}{2}}v, -v) - v_6(u + {\frac{3}{2}}v, -v).   
\label{eq:16}
\end{eqnarray}
As one can check, the expansions (10), (11), and (12) do really 
satisfy these relations. Moreover, consideration of all the remaining 
limiting cases (see, e. g. Refs.\cite{MSS,CPV}) shows that these 
RG series are in accord with their counterparts obtained earlier 
for the Ising \cite{GZ2,SOU} and O(n)-symmetric \cite{SOUK} models. 

Let us use the expansions just found for estimation of the universal 
critical values of the sextic effective coupling constants for the cubic 
ferromagnets, i. e., in the physical case $n = 3$. To obtain 
the numbers of interest from the divergent (asymptotic) RG series, 
a proper resummation procedure should be applied. Here we use the 
Pad\'e-Borel-Leroy resummation technique which demonstrates high 
numerical effectiveness both for simple [Ising and O(n)-symmetric] 
models \cite{S,SOUK,BNM,AS} and for complicated anisotropic systems 
preserving their internal symmetries (see, e. g. Ref.\cite{AS1} for 
detail). Since the expansions of quantities depending on two variables 
$u$ and $v$ are dealt with, the Borel-Leroy transformation is taken 
in a generalized form: 
\begin{equation}
f(u,v)=\sum_{ij}c_{ij}u^iv^j=\int\limits_0^\infty
e^{-t} t^b F(ut,vt)dt, \qquad
F(x,y)=\sum_{ij}{\frac{c_{ij}x^i y^j}{{(i+j+b)!}}}. 
\label{eq:17}
\end{equation}
To perform an analytical continuation, the resolvent series
\begin{equation}
\tilde F(x,y,\lambda )=\sum_{n=0}^\infty \lambda
^n\sum_{l=0}^n\sum_{m=0}^{n-l}{\frac{c_{l,n-l}x^ly^{n-l}}{{n!}}}
\label{eq:18}
\end{equation}
is constructed which is a series in powers of $\lambda$ with the
coefficients being uniform polynomials in $u$, $v$ and then Pad\'e
approximants $[L/M]$ in $\lambda $ at $\lambda = 1$ are used. 

With the four-loop RG expansions in hand, we can construct, 
in principle, three different Pad\'e approximants: [2/1], [1/2], 
and [0/3]. To obtain proper approximation schemes, however, only 
diagonal [L/L] and near-diagonal Pad\'e approximants should be employed 
\cite{BGM}. That is why further, when estimating $u_6^*$, $q_6^*$, and 
$v_6^*$, we limit ourselves to the approximants [2/1] and [1/2]. Moreover, 
the diagonal Pad\'e approximant [1/1] will be also dealt with although 
this corresponds, in fact, to the use of the lower-order, three-loop 
RG approximation. 

The algorithm for estimating of the universal critical values of sextic 
effective couplings is as follows. Since, in fact, we have 
the power series for the ratios $R_u = u_6/u^2$, $R_q = q_6/(u v)$, and
$R_v = v_6/v^2$ rather than for $u_6$, $q_6$, and $v_6$ themselves, we 
work with the RG series for these ratios. They are resummed in three 
different ways based on the Borel-Leroy transformation and the Pad\'e 
approximants just mentioned. The Borel-Leroy integral is evaluated as 
a function of the shift parameter $b$ under $u = u^*$, $v = v^*$. For 
the cubic fixed-point coordinates the values given by the six-loop RG 
analysis at $n = 3$ are adopted \cite{CPV}: $u^* = 1.321$, $v^* = 0.117$.  
The optimal value of $b$ providing the fastest convergence of the 
iteration scheme is then determined. It is deduced from the condition 
that the Pad\'e approximants employed should give, for $b = b_{opt}$, 
the values of $R_u$ ($R_q$, $R_v$) which are as close as possible to 
each other. Finally, the average over three estimates for $R_u$ 
($R_q$, $R_v$) is found and claimed to be a numerical value of this 
universal ratio. 
 
To demonstrate how such a procedure works, in Table 1 the results 
of the corresponding calculations are presented. The empty cells 
in this table reflect the fact that for some values of 
the shift parameter $b$ Pad\'e approximant [1/2] turns out to be 
spoiled by the "dangerous" poles, i. e., by the poles at positive or 
small negative $t$. As one can see, for $u_6^*$ (an integer) $b_{opt}$, 
providing a maximal closeness of the estimates given by three 
working Pad\'e approximants, is equal to 2, while for $q_6^*$ and 
$v_6^*$, $b_{opt} = 3$. So the results of our four-loop RG analysis 
are as follows:
\begin{equation}
u_6^* = 0.842, \qquad 
q_6^* = 0.175, \qquad 
v_6^* = 0.0108. 
\label{eq:19} \\
\end{equation}

How close to the exact universal values of sextic couplings may these 
numbers be? Obviously, the accuracy of our estimates is limited both 
by the accuracy achieved in the course of evaluating the cubic 
fixed-point coordinates $u^*$, $v^*$ themselves and by the speed of 
convergence of the iteration procedure employed. The influence of the 
latter factor may be characterized by the sensitivity 
of the numerical results with respect to variation of $b$. As is seen 
from Table 1, the  values of $u_6^*$, $q_6^*$, and $v_6^*$, averaged 
over the working Pad\'e approximants, vary by no more than $\pm 0.005$, 
$\pm 0.002$, and $\pm 0.0001$, respectively, when $b$ runs from 0 to 
20. On the other hand, these error bars are too small to be referred 
to as realistic ones. Their conservative counterparts may be deduced, 
accepting that the true universal values of renormalized sextic 
couplings should lie within the intervals of the variation of the 
estimates given by the leading Pad\'e approximant, namely, by the 
higher-order pole-free approximant [2/1]. These intervals are easily 
extracted from Table 1 and are as follows: $\pm 0.03 ~(u_6^*)$, 
$\pm 0.01 ~(q_6^*)$, $\pm 0.001 ~(v_6^*)$. Contrary to the 
aforementioned tiny ranges, the error bars thus determined look quite 
realistic. Moreover, they are found to be large enough to cover the 
inaccuracy produced by the uncertainty of the cubic fixed-point 
location. Hence, we adopt these error bars as the final ones. 

With the universal critical values of the quartic and sextic couplings 
in hand, we are able to estimate the reduced anisotropy of the 
sixth-order nonlinear susceptibility. Combining Eqs. (4), (5), and (19), 
we obtain  
\begin{equation}
\delta^{(6)} = 0.102 \pm 0.02. 
\label{eq:20} \\
\end{equation}

\section{Eighth-order nonlinear susceptibility and discussion}
\label{sec:4}

Let us discuss the role played by the sextic coupling constants 
in forming the magnitudes of $\chi_{c}^{(6)}$, $\chi_{d}^{(6)}$, and 
$\delta^{(6)}$. It may be shown that, in fact, their contrubutions to 
the sixth-order susceptibility and the anisotropy parameter are very 
small. Indeed, neglecting $u_6$, $q_6$, and $v_6$ in Eq. (4) changes 
the critical value of $\delta^{(6)}$ by about 4\%. In other words, 
the anisotropy of $\chi^{(6)}$ near $T_c$ is fixed practically by the 
quartic coupling constants only. A similar situation should take place 
for the higher-order (eighth-order, tenth-order, etc.) susceptibilities. 
To demonstrate this, we consider the eighth-order susceptibility 
$\chi^{(8)}$ for the isotropic (Heisenberg) model. As $\chi^{(4)}$ and 
$\chi^{(6)}$, it may expressed via the effective coupling constants    
\begin{equation}
\chi^{(8)} = - 40320 {\frac{\chi^4}{m^9}} \Bigl( 
96 u_4^3 - 24 u_4 u_6 + u_8 \Bigr).
\label{eq:21} 
\end{equation}
Up to now, the universal critical value of the octic coupling constant 
was determined using the field-theoretical RG approach in three 
dimensions \cite{SOUK}, the biased $\epsilon$ expansion technique 
\cite{PV}, and the exact renormalization group machinery \cite{TW}. The 
methods employed lead to the estimates $u_8^* = 0.168$, $u_8^* = 0.36$, 
and $u_8^* = 0.145$, respectively. Although these numbers are 
considerably scattered, all of them are several times smaller than 
$u_4^* = 0.794$ (Ref.\cite{GZ1}) and $u_6^* = 0.951$ (Ref.\cite{SOUK}). 
It means that, because of the big numerical coefficients of the first two 
terms in brackets of Eq. (21), the contribution of $u_8$ to $\chi^{(8)}$ 
is negligible at criticality. Since the cubic fixed point is located very 
near the Heisenberg one at the RG flow diagram, the same conclusion is 
valid for the 3D cubic model. This enables us to estimate 
the critical anisotropy of $\chi^{(8)}$ without the calculation of the 
universal values of the octic coupling constants. Making in Eq. (21) 
the substitutions $u_4 \to u_4 + v_4$, $u_6 \to u_6 + q_6 + v_6$ and 
$u_4 \to u_4 + v_4/3$, $u_6 \to u_6 + q_6/3 + v_6/9$ with $u_8$ omitted, 
we obtain expressions for $\chi_c^{(8)}$ and $\chi_d^{(8)}$, 
respectively. Using then the known coordinates of the cubic fixed point 
and the critical values of the sextic coupling constants (19), we obtain 
the reduced anisotropy of the eighth-order susceptibility: 
\begin{equation}
\delta^{(8)} =  {\frac{|\chi_{c}^{(8)} - \chi_{d}^{(8)}|} 
{\chi_{c}^{(8)}}} = 0.144 \pm 0.04. 
\label{eq:22} \\
\end{equation}

The estimates (6), (20), (22) for $\delta^{(4)}$, $\delta^{(6)}$,  
and $\delta^{(8)}$ at criticality do not look too small to prevent 
the detection of the universal cubic anisotropy in physical and 
computer experiments or by means of a thorough analysis of the 
high-temperature expansions \cite{BC,CPRV}. It is well known, however, 
that the approach of the universal critical regime is controlled by the 
tiny exponent $\omega \approx 0.01$ \cite{CPV}, making both an 
experimental study and simulation of the true mode of critical behavior 
rather difficult \cite{CH}. We believe that the results obtained in 
this paper will help those searching for the cubic asymptotic regimes 
in real systems and lattice models to choose a proper strategy and the 
quantities to be measured.  

\section{Conclusion}
\label{sec:5}

To summarize, we have evaluated the nonlinear susceptibilities of the 
forth, sixth, and eighth orders for the 3D cubic ferromagnet and estimated 
the parameters $\delta^{(4)}$, $\delta^{(6)}$, and $\delta^{(8)}$ 
characterizing their anisotropy at the critical point. In the course 
of this study, we have calculated the four-loop RG expansions for 
the effective coupling constants $u_6$, $q_6$, and $v_6$ entering the 
small-magnetization free-energy expansion and found, resumming the RG 
series by the Pad\'e-Borel-Leroy technique, the universal critical 
values of the renormalized sextic couplings at the cubic fixed point. 
The values of the anisotropy parameters $\delta^{(i)}$ have turned out 
to be appreciable at criticality, indicating that the cubic 
(non-Heisenberg) critical behavior should be, in principle, detectable 
in physical and computer experiments.  
 
\acknowledgments

This work was supported by the Russian Foundation for Basic Research 
(Grant No. 01-02-17048), by the Ministry of Education of Russian 
Federation via the Grant Center of Natural Sciences (Grant No. 
E00-3.2-132), and by the Federal Program "Integratsiya" (Project No. 
A 0150). One of the authors (A. I. S.) gratefully acknowledges 
also the support of the International Science Foundation via 
Grant p00--1493.

\narrowtext
\begin{table}
\caption{Numerical estimates for the universal values of sextic 
effective coupling constants $u_6$, $q_6$, and $v_6$ obtained from 
the four-loop RG expansions (10), (11), and (12) 
resummed by the Pad\'e-Borel-Leroy technique using approximants 
$[2/1]$, $[1/2]$, and $[1/1]$. The empty cells are due to the 
"dangerous" poles spoiling the corresponding Pad\'e approximants.}
\begin{tabular}{|cc|ccccccc|}
b &  & 0 & 1 & 2 & 3 & 5 & 10 & 20 \\
\hline
$u_6^*$ & [2/1] & 0.8654 & 0.8514 & 0.8427 & 0.8368 
& 0.8294 & 0.8206 & 0.8144 \\
& [1/2] & 0.8345 & - & 0.8430 & 0.8384 & 0.8356 & 0.8328 & 0.8309 \\
& [1/1] & 0.8097 & 0.8295 & 0.8400 & 0.8466 & 0.8543 & 0.8625 & 0.8679 \\
\hline
$q_6^*$ & [2/1] & 0.1834 & 0.1791 & 0.1765 & 0.1747 
& 0.1725 & 0.1699 & 0.1680 \\
& [1/2] & 0.1729 & 0.1714 & - & 0.1738 & 0.1737 & 0.1729 & 0.1723 \\
& [1/1] & 0.1653 & 0.1707 & 0.1736 & 0.1754 & 0.1775 & 0.1797 & 0.1812 \\
\hline
$v_6^*$ & [2/1] & 0.01168 & 0.01128 & 0.01103 & 0.01086 
& 0.01065 & 0.01041 & 0.01024 \\
& [1/2] & 0.01062 & 0.01059 & - & 0.01090 & 0.01072 & 
0.01064 & 0.01060 \\
& [1/1] & 0.00989 & 0.01035 & 0.01060 & 0.01075 & 0.01093 & 0.01111 
& 0.01124 \\
\tableline
\end{tabular}
\label{table2}
\end{table}
\end{document}